\documentclass[prl,aps,floatfix,twocolumn,preprintnumbers]{revtex4-2}%
\usepackage[normalem]{ulem}
\usepackage{epsfig}
\usepackage{amsmath}
\usepackage{amsfonts}
\usepackage{amssymb}
\usepackage{xcolor}
\usepackage[caption=false]{subfig}
\usepackage{graphicx}%
\providecommand{\U}[1]{\protect\rule{.1in}{.1in}}
\providecommand{\U}[1]{\protect\rule{.1in}{.1in}}

\begin{document}
\preprint{ }
\title{Suppressed fluctuations as the origin of the static magnetic order in strained Sr$_{2}%
$RuO$_{4}$}
\author{Bongjae Kim$^{1}$}
\email{bongjae.kim@kunsan.ac.kr}
\author{Sergii Khmelevskyi$^{2}$}
\author{Cesare Franchini$^{3,4}$}
\author{I. I. Mazin$^{5,6}$}
\email{imazin2@gmu.edu}
\affiliation{$^{1}$ Department of Physics, Kunsan National University, Gunsan, 54150, Korea}
\affiliation{$^{2}$ Center for Computational Materials Science, Institute for Applied
Physics, Vienna University of Technology, Wiedner Hauptstrasse $8$ - $10$,
$1040$ Vienna, Austria}
\affiliation{$^{3}$ University of Vienna, Faculty of Physics and Center for Computational
Materials Science, Vienna A-1090, Austria}
\affiliation{$^{4}$ Dipartimento di Fisica e Astronomia, Universit\`{a} di Bologna, 40127
Bologna, Italy}
\affiliation{$^{5}$ Department of Physics and Astronomy, George Mason University, Fairfax,
VA 22030}
\affiliation{$^{6}$ Quantum Science and Engineering Center, George Mason University,
Fairfax, VA 22030}
\date[Dated: ]{\today}

\begin{abstract}
Combining first principle density functional calculations and Moriya's self-consistent renormalization theory, we explain the recently reported counterintuitive appearance of an ordered magnetic state in uniaxially strained Sr$_{2}$RuO$_{4}$ beyond the Lifshits transition.
We show that strain weakens the quantum spin ﬂuctuations, which destroy the static order, more strongly than the tendency to magnetism. A  different rate of decrease of the spin fluctuations vs. magnetic stabilization energy promotes the onset of a static magnetic order beyond a critical strain.
\end{abstract}
\maketitle

\emph{Introduction. }
After many years of pursuing triplet superconductivity in Sr$_{2}$RuO$_{4}$, recent studies provide evidence in favor of a singlet state~\cite{Chronister2021,Petsch2020}.
However, while most theorists believe that spin fluctuations are important for superconductivity, there is no consensus on the detailed mechanism or the pairing symmetry. Some of the proposals include $d_{x^{2}-y^{2}}%
$~\cite{Hassinger2017,Steppke2017,Gingras2022}, $s +id_{xy}$~\cite{Roemer2021}, $d_{xz}+id_{yz}%
$~\cite{Zutic2005,Suh2020}, or $g_{xy(x^{2}-y^{2})}$~\cite{Gingras2022,Kivelson2020}.
Experiments under uniaxial strain have been instrumental in the quest for elucidating the pairing symmetry~\cite{Hicks2014,Steppke2017}. One of the most impressive pieces of evidence was the observation of a maximum in $T_{c}$ and $H_{C2}$ at the critical strain corresponding to the Lifshitz
transition~\cite{Steppke2017,Luo2019}. In systems where the Fermi level ($E_F$) lies close to a Van Hove singularity (VHs), small perturbations can induce a Lifshitz transition involving a tuned shift of the Van Hove point across $E_F$ associated with a change of the  Fermi surface topology~\cite{Lifshitz1960,Barber2018,Sunko2019}. The latest addition in this direction is establishing an ordered static magnetic state under strain beyond the Lifshitz transition~\cite{Grinenko2021}.

This result is counterintuitive because the magnetism is nearly universally suppressed by pressure as the atomic distances, $a$, become smaller (See Refs~\cite{Shimizu2001,Cohen1997} for representative examples): the bandwidth scales as $1/a^2$ and DOS as $a^2$. Hence, $I(0)N(0)$, which determines the tendency towards  ferromagnetism in Stoner picture, becomes smaller and enhanced itinerancy eventually wins over the spin exchange splitting. Here, $I(0)$ and $N(0)$ are Stoner factor and density of states at Fermi level, respectively. Understanding this paradox may bring up new and novel progress in understanding the physics of this compound, which has a nontrivial magnetic energy landscape~\cite{BKim2022}. The emergence of magnetism under pressure, as stated, is very uncommon and usually associated with materials close to the itinerant magnetism, where the long-range order is suppressed by the fluctuations (cf. Fe-based superconductors, Ref.~\cite{Khasanov2020}). Of course, this is the average behavior and around the Lifshitz transition, this picture is not valid. Further, the leading instability in Sr$_{2}$RuO$_{4}$ is not ferromagnetic but at an incommensurate $\mathbf{q}$, where the susceptibility is not affected by the Lifshitz transition.

Such anomalous pressure effect on magnetism can be understood within Moriya's
self-consistent renormalization (SCR) theory~\cite{Moriya}. It stipulates that
the magnetization in itinerant magnets is soft and fluctuates in amplitude.
Assuming Gaussian fluctuations with the mean square amplitude, $\xi,$ it was
shown that if the total energy is expanded with the magnetic order parameter
$M$ as
\begin{equation}
E=a+bM^{2}+cM^{4}+dM^{6}+...,\label{1}.%
\end{equation}
According to the fluctuation-renormalized expansion, the corresponding
coefficients change as%
\begin{align}
\tilde{b} &  =b+\frac{5}{3}c\xi^{2}+\frac{35}{9}d\xi^{4}+...\nonumber\\
\tilde{c} &  =c+\frac{14}{3}d\xi^{2}+....\label{2}%
\end{align}
and so on.
Obviously, this increases the value of the inverse spin
susceptibility, $2b,$ and its sign changes from negative to positive for
$\xi^{2}\gtrsim|3b/5c|$. Hence, the spin fluctuation in an itinerant magnet can lead to \emph{(i)} the reduction of spin susceptibility in paramagnetic materials, \emph{(ii)} the decrease of the average magnetic moment, or
\emph{(iii)} the suppression of the long-range magnetic order in a system
where mean-field theories, such as density functional theory (DFT), predicts to be an ordered magnet.

SCR theory has been successfully applied to various (near-)ferromagnetic
materials, such as Pd~\cite{Larson2004}, Ni$_{3}$Ga and Ni$_{3}$%
Al~\cite{Aguayo2004}, ZrZn$_{2}$~\cite{Mazin2004}, and notable other systems where the SCR theory is somewhat more straightforward.
Also, it was called up in connection with Fe-based superconductors~\cite{Mazin2008,Ortenzi2015}. Moriya, in his
book~\cite{Moriya}, emphasizes that, while the entire frequency and momentum dependence of spin susceptibility, \textit{via} the fluctuation-dissipation theorem, determines the value of $\xi,$ a significant role is played by the phase space, $i.e.$, the fraction of the Brillouin zone (BZ) that is close to a magnetic instability.

This point was clearly demonstrated in Ref.~\cite{Aguayo2004}, where
$\xi$ was estimated from first principle calculations for two similar
compounds, Ni$_{3}$Al and Ni$_{3}$Ga. Within DFT, both systems are ferromagnetic, but, in reality, only the former is, while Ni$_{3}$Ga is a strongly renormalized paramagnet. Counterintuitively, the calculated magnetic moment and magnetic stabilization energy were higher in Ni$_{3}$Ga. This apparent paradox was resolved in Ref.~\cite{Aguayo2004} by the fact that the instability in Ni$_{3}$Ga, while stronger, is also considerably less localized in the momentum space (the characteristic volume of the unstable part of the BZ is more than twice larger), which leads to a larger $\xi,$ and a stronger suppression. As a result, from the view of mean-field theory, the magnetic instability is entirely suppressed in the $more$ ferromagnetic material, while in the $less$ one, the magnetism survives~\cite{Aguayo2004}.

In this Letter, we argue that Sr$_{2}$RuO$_{4}$ represents a similar case with the spin-density wave (SDW) type instability: The tendency toward SDW-antiferromagnetism is stronger for the unstrained material, but the static order is not established due to the even more substantial spin fluctuations. We show that when uniaxial stress is applied, the fluctuations are suppressed more strongly than the tendency towards magnetism which causes the emergence of a static magnetic phase.

\emph{Methods. }
We employed the Vienna \emph{ab initio} simulation package
(VASP)~\cite{Kresse1993,Kresse1996} within the  Generalized gradient approximation (GGA) by Perdew-Burke-Ernzerhof functional~\cite{Perdew1996}. The energy cut for the plane waves of 600 eV was
used with a Monkhorst-Pack $k$-mesh of 17 $\times$ 17 $\times$ 10 for the
primitive unit cell. For the spin spiral calculations with various
$\mathbf{q}$-values, we employed the generalized Bloch
theorem~\cite{Sandratskii1991}, allowing spiral calculations for an
arbitrary wave vector without using the supercells.
We have further used a very stringent convergence
criteria of $10^{-7}$ eV in most cases.

\begin{figure}[t]
\centering
\includegraphics[width=.95\columnwidth]{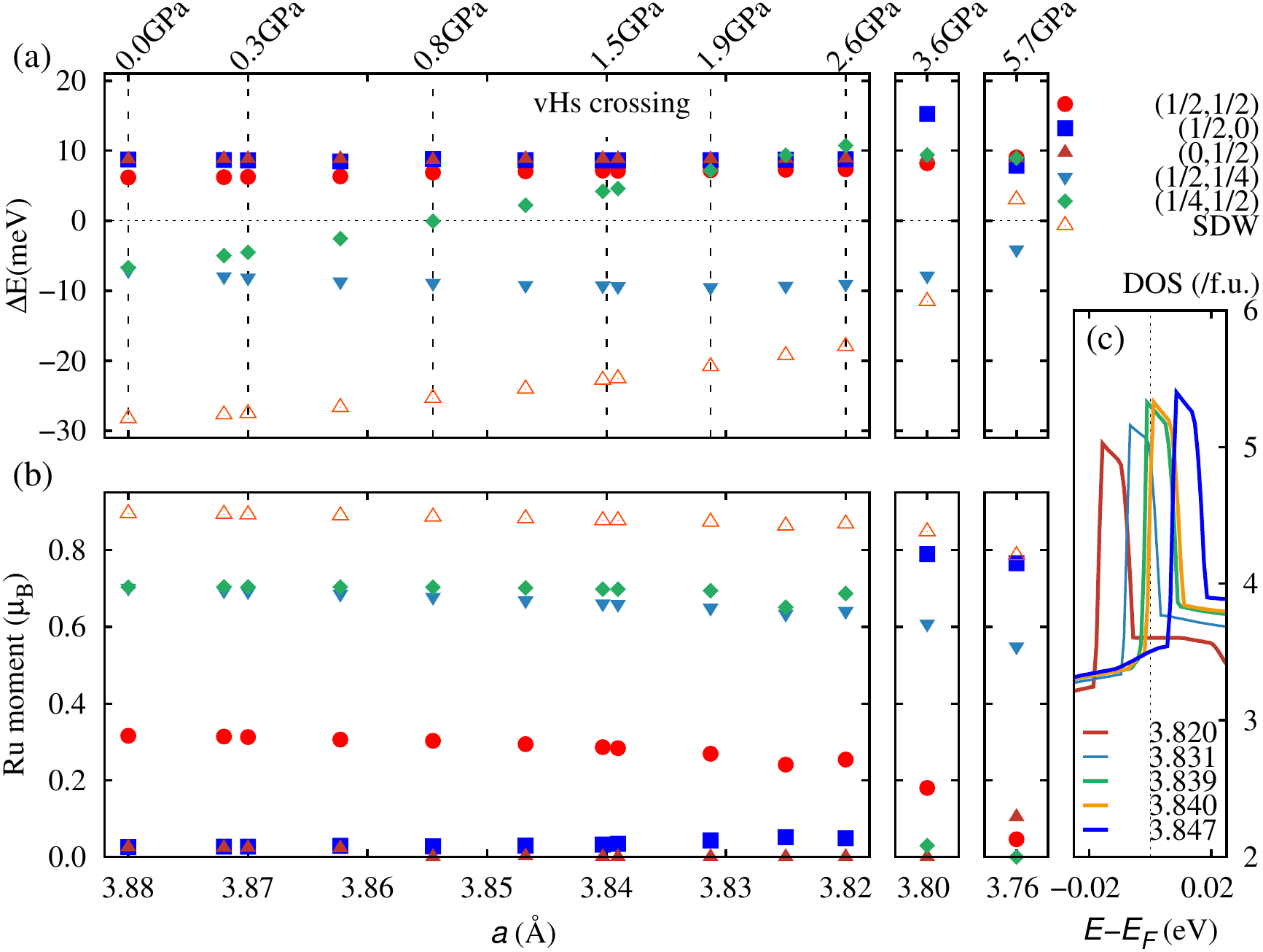}\caption{ Uniaxial strain
dependent evolution of (a) total energy and (b) Ru local magnetic moment of
Sr$_{2}$RuO$_{4}$ for various magnetic $\mathbf{q}$ ordering vectors. The calculated SDW wave vector is $\mathbf{q}_{SDW}=(0.29,0.29,0)$. For the total energy evaluation, we
set $\Delta E$ as the energy difference of each configuration with
nonmagnetic one. The compressively strained $a$ lattice parameter and
corresponding stress is indicated in the $x$-axis of the plot. (c) Position of
the VHs as a function of strain. For a critical strain at 3.840 \AA,
corresponding to a pressure of 1.5GPa (see panel a and b), the VHs crosses $E_F$ establishing
a Lifshitz transition.}%
\label{fig1}%
\end{figure}

To obtain a series of strained structures, we fixed the $a$ lattice
parameter and fully optimized $b$ and $c$ parameters and the internal
positions. Due to the well-known fact that GGA overestimates the equilibrium
volume at a given pressure, the critical stress (but not the critical strain)
in our calculations is likely overestimated (we get the Lifshits transition at
$\sigma=1.5$ GPa, see Fig.~\ref{fig1}, about twice the value estimated in Ref.
\cite{Grinenko2021}), but the important part is that we trace the evolution of
magnetic properties well past this transition.

\emph{Results. }
Sr$_{2}$RuO$_{4}$ does not order magnetically down to low
temperatures; however, it was predicted from DFT calculations~\cite{Mazin1999} and later confirmed by the neutron diffraction that it features strong spin fluctuations with the wave vector $\mathbf{q}\sim(0.3,0.3,0)$~\cite{Sidis1999, Iida2011} (here and below, we give the wave vectors in the orthorhombic reciprocal lattice units, $i.e.,$ $2\pi/a$). This incommensurate SDW feature, from the nesting of 1D $\alpha$ and $\beta$ Fermi surfaces, can be well-captured by DFT in the unstrained case using approximate commensurate $\mathbf{q}$~\cite{BKim2017}. While we did not attempt to locate the exact position of the incommensurate SDW instability under strain, we could monitor its evolution using our commensurate $\mathbf{q}$-space grid. In our case, we obtained $\mathbf{q}_{SDW}$ at $(0.29,0.29,0)$ as the ground state for the unstrained case, which gives excellent agreement with the neutron diffraction study (from now on, we will omit $q_{z}=0$ for brevity).

\begin{figure*}[t]
\centering
\includegraphics[width=\textwidth]{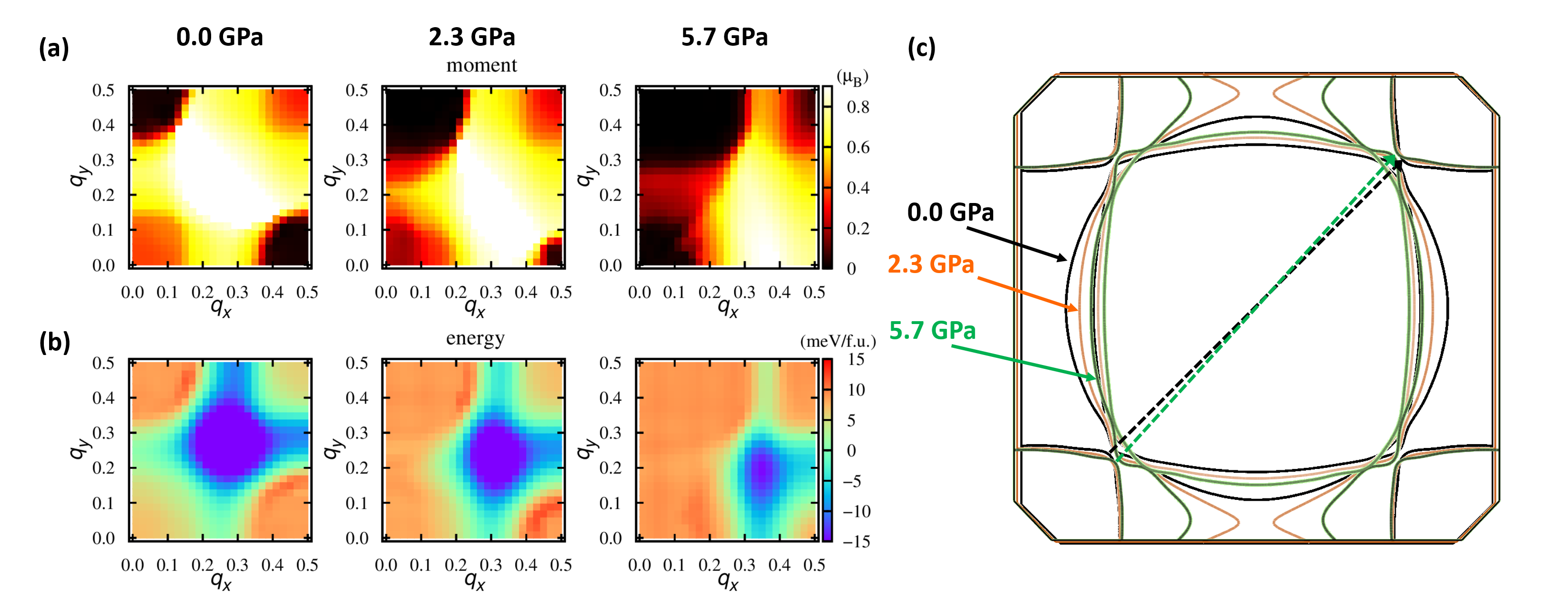}\caption{(a) Ru local magnetic
moment and (b) energy calculation for ($\mathbf{q_{x}}$,$\mathbf{q_{y}}$)
plane for three representative stress cases: unstrained, 2.3 GPa and 5.7 GPa.
2.3 GPa corresponds to the case beyond the Lifshitz transition, and 5.7 GPa to the
highly strained limit. Here, the $\mathbf{q}$-grid calculation is performed
with varying $\mathbf{q_{x}}$ and $\mathbf{q_{y}}$ values with the interval of
$\Delta\mathbf{q} = 0.02 \frac{2\pi}{a}$ at $\mathbf{q_{z}}=0$. (c) The corresponding Fermi surface plots. The diagonal dashed arrows indicate the nesting vectors for unstrained and 5.7 GPa
cases. }%
\label{fig2}%
\end{figure*}

First, to inspect the correlation between the leading magnetic instability and uniaxial pressure in  Sr$_{2}$RuO$_{4}$, we have calculated the total energy and Ru local magnetic moment for representative magnetic orderings in a wide range of stress levels and pressures. In addition to the $\mathbf{q}_{SDW}$ phase, we have considered the $\mathbf{q}=(1/2,1/4)$ order suggested by R{\o}mer $et$ $al$~\cite{Roemer2020} and the checkerboard ordering $\mathbf{q}=(1/2,1/2)$, as well as  $\mathbf{q}=(1/2,0)$ and $(0,1/2)$ configurations~\cite{BKim2017}, based on the observation that only the first nearest neighbor exchange responds to uniaxial stress~\cite{BKim2017}.
In Fig.~\ref{fig1}, we display the energy difference of each configuration with the nonmagnetic one ($\Delta E$) and Ru local moment for uniaxially strained Sr$_{2}$RuO$_{4}$. We find that the SDW phase is the most stable state up to 3.6 GPa. As shown in Fig~\ref{fig1}(c), the VHs crosses the Fermi energy at around 1.5 GPa, and the SDW phase remains the ground state well beyond that. Uniaxial strain breaks the $C_{4}$ symmetry and bifurcates the first nearest neighbor exchange interactions between Ru ions~\cite{BKim2017}. Hence, $\mathbf{q}=(1/4,1/2)$ and $\mathbf{q}=(1/2,1/4)$ are progressively split in energy. We see that at the highly-strained limit of 5.7 GPa, $\mathbf{q}=(1/2,1/4)$ is lower than $\mathbf{q}_{SDW}$ in energy, which, however, does not indicate the
$\mathbf{q}=(1/2,1/4)$ is the ground state, as discussed in detail later.

Interestingly, the size of the Ru moment strongly varies for different
$\mathbf{q}$ values. The $\mathbf{q}_{SDW}$ moment is the largest for all
studied ranges, while the one for $\mathbf{q}=(1/2,0)$ is negligibly small. The sizes of the moment for $\mathbf{q}=(1/4,1/2)$ and
$\mathbf{q}=(1/2,1/4)$ also bifurcate, and, at 3.6 GPa, the moment suddenly vanishes for $\mathbf{q}=(1/4,1/2)$.

For the unstrained case, while strong $\mathbf{q}_{SDW}$ tendency exists, the long-range magnetic order does not set in due to the strong spin-fluctuations.
As the system undergoes uniaxial strain, $\mu$SR experiments have found the stabilization of a magnetic order beyond the Lifshitz transition
point~\cite{Grinenko2021}. However, the actual magnetic pattern is unknown, albeit an incommensurate SDW was suggested~\cite{Grinenko2021}. On the other hand,
preliminary nuclear magnetic resonance (NMR) data, while also suggestive with magnetic order, do not see expected broadening at low
temperatures~\cite{Stuart}.

Our calculations, consistently with $\mu$SR, show that the DFT (that is to
say, mean field) ground state is always an incommensurate SDW, even though its
wave vector slowly shifts from $(0.29,0.29)$ toward $\mathbf{q}=(1/2,1/4)$
[Fig.~\ref{fig2}(b)]. As expected, while SDW is always stable at the mean
field level, the stabilization energy [Fig.~\ref{fig2}(b)] and the equilibrium
magnetic moment [Fig.~\ref{fig2}(a)] are considerably reduced by the strain.

The next step is to estimate the effect of the fluctuations. For materials
close to ferromagnetism, first principle calculations of the $\xi$ parameter
(Eq. \ref{2}) are tedious but feasible~\cite{Larson2004, Aguayo2004}. For those close to antiferromagnetism, we do not have a similar recipe.  Therefore, we have adapted \textquotedblleft a poor man's\textquotedblright\ approach, estimating
$\xi^{2}$ as $\left\langle M^{2}(\mathbf{q})\right\rangle ,$
which is obtained by averaging the squared Ru local magnetic moment over
the entire Brillouin zone.
Ideally, knowing the frequency- and momentum-dependent spin susceptibility, $\chi(\mathbf{q},\omega)$, one could evaluate $\xi^{2}$~\cite{Moriya}. Lacking this information, we assume that the frequency dependence does not affect the trend (similarly, the $\omega$ dependence was neglected, for instance, in Ref.~\cite{Kubler}, where $\xi^{2}$ was estimated for some ferromagnets in the high-temperature limit). Since we, on the contrary, are interested in zero-point fluctuation, we can further assume that fluctuations at any $\mathbf{q}$ are excited with the same probability and the amplitude proportional to $M(\mathbf{q})$, from where our \textquotedblleft poor man\textquotedblright\ formula follows. While this estimate may be quantitatively inexact, it should provide us with the correct trend under strain.

To this end, we have calculated, for three stresses of 0.0 GPa, 2.3 GPa, and 5.7 GPa, the moments and magnetic stabilization energies on a full 2D-grid in $\mathbf{q_{x}}$ and
$\mathbf{q_{y}}$, while keeping $\mathbf{q_{z}}=0$. The results are shown in Fig.~\ref{fig2} (a) and (b). For the unstrained case, a large moment  ($>0.8\mu_{B}$) can be found in a broad region centered at $\mathbf{q}_{SDW}=(0.29,0.29)$. The size of the moment strongly depends on $\mathbf{q}$, and almost vanishes close to $\mathbf{q}=(1/2,0)$ and $\mathbf{q}=(0,1/2)$.
The large \textbf{q} range where a strong instability occurs demonstrates
the highly fluctuating nature of the spin moment in Sr$_{2}$RuO$_{4}$.
As the uniaxial stress is imposed, the lowest energy position is gradually
moved asymmetrically to $\mathbf{q}=(0.32,0.24)$, for 2.3 GPa, and $\mathbf{q}=(0.34,0.20)$ for 5.7 GPa (See Fig.~\ref{fig2}(b)). According to our calculation, the magnetic instability is of SDW-type, not the commensurate one $\mathbf{q}=(1/2,1/4)$ suggested in Ref.~\cite{Roemer2020} and not checkerboard $\mathbf{q}=(1/2,1/2)$. While there is an apparent tendency of the ground-state $\mathbf{q}$ to shift from the original SDW one, $\mathbf{q}_{SDW}=(0.29,0.29)$, toward $\mathbf{q}=(1/2,1/4)$, the magnetism of the system remains incommensurate.
Our calculations show that, while proximity to an SDW instability
in unstrained Sr$_{2}$RuO$_{4}$ is definitely related to nesting~\cite{Mazin2004},
the exact position is not given by the nesting vector (which is related to the
imaginary, not the real part of susceptibility\cite{CDW}), and in fact the
evolution of the SDW vector with strain cannot be attributed to the evolution
of nesting (Fig. \ref{fig2})~\cite{suppl}.

Estimating $\xi^{2}\approx\left\langle M^{2}(\mathbf{q)}\right\rangle$, we
obtain for 0.0 GPa, 2.3 GPa and 5.7 GPa stress, respectively, 0.43, 0.36, and
0.22 ${\mu_{B}}^{2}$, demonstrating the progressive suppression of the spin
fluctuations. The apparent shrinking of the area where the moment survives
(Fig.~\ref{fig2}(a)) demonstrates the dramatic reduction of the phase space
available for fluctuations under strain.

 While the leading instability has, of course, its origin in the momentum space, it is instructive to look at it also from the real space point of view. As discussed in our previous paper~\cite{BKim2017}, the Fermi-surface driven instability, when mapped onto the nearest-neighbor Heisenberg
Hamiltonian, results in three sizeable exchange parameters, corresponding to the ferromagnetic coupling along the 110 bond, $J_{\rm 110}$ and two antiferromagnetic couplings, along the 100 and 200 bonds, so that $|J_{\rm 200}|>|J_{\rm 110}|>|J_{\rm 100}|$. Note that single $J_{\rm 100}$ generate a ${\bf q}=(1/2,0)$ order, and $J_{\rm 200}$ a ${\bf q}=(1/4,0)$ order, \emph{etc}. In Ref. ~\cite{BKim2017} we estimated, using the real-space Disordered Local Moments method, the
effect of uniaxial strain (without reoptimizing the atomic position) on the parameters $J$s, and found that only $J_{\rm 100}$ is affected, by splitting into
different  $J_{\rm 100}$ and $J_{\rm 010}$. For sufficiently large strain, $J_{\rm 100}$ dictates the antiferromagnetic order along $x$, that is, $q_x=1/2$, while (still the largest) $J_{\rm 200}$ forces $q_y$ to get close to 1/4, thus promoting the $(1/2,1/4)$ order. However, as discussed above, direct calculations
show that this limit is never achieved in the considered stress range, albeit the leading instability shifts from (1/3,1/3) to this general direction. Thus, the evolution of the instability {\bf q} vector can be roughly described as the competition between the Fermi surface driven instability and the nearest neighbor superexchange.

\begin{figure}[t]
\centering
\includegraphics[width=.95\columnwidth]{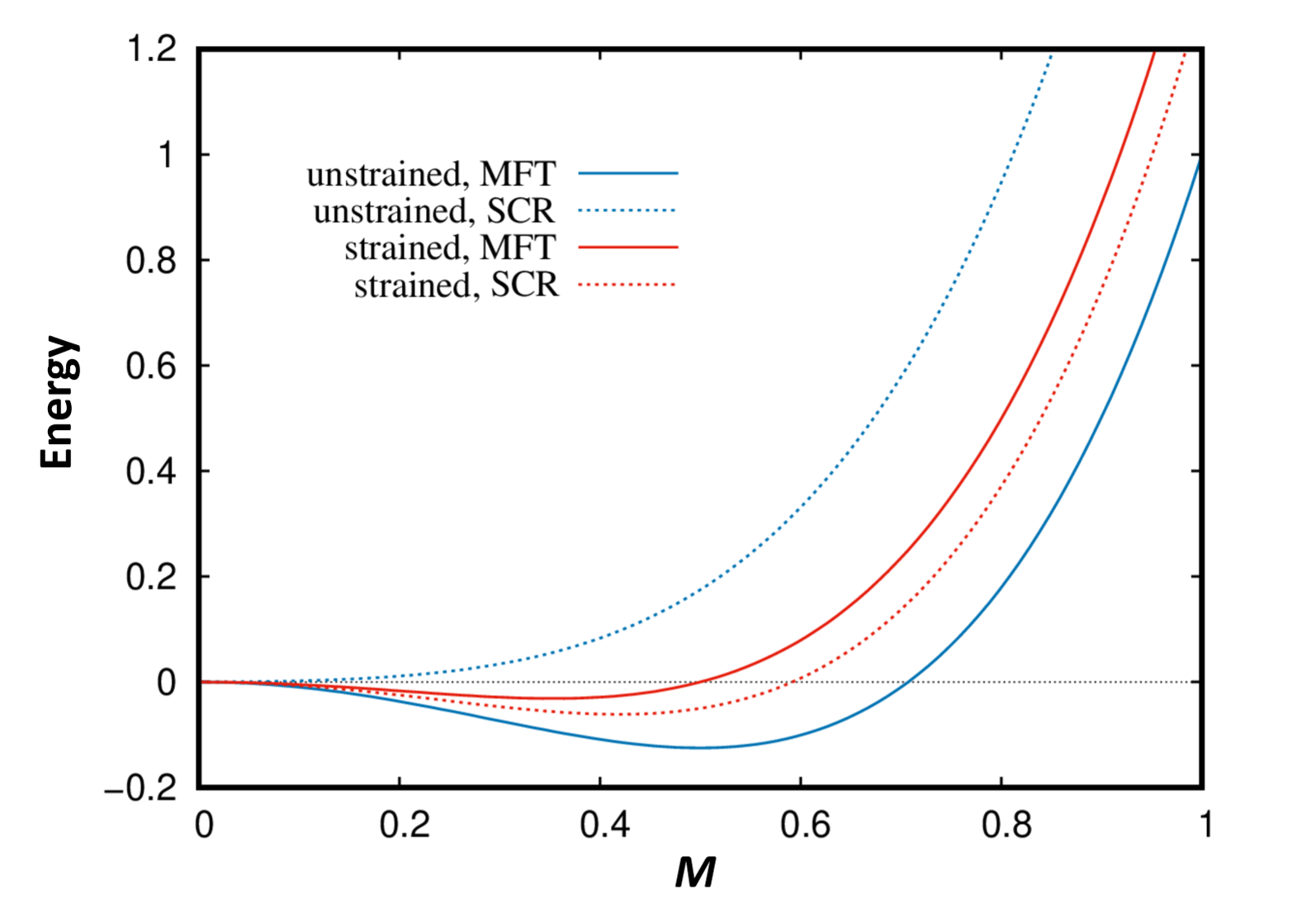}\caption{
A toy model illustrating that within the SCR theory, the overall suppression of magnetism may, paradoxically, lead to the establishment of long-range magnetic order. The figure shows a model magnetic Hamiltonian, that is, the energy as a function of a magnetic order parameter $M$.
The mean-field theory, in this particular model and range of parameters (see the main text), yields a magnetic ground state for both unstrained and strained cases (solid lines). On the other hand, with SCR theory, where the spin-fluctuations are included per Eq. \ref{2}, one obtains a magnetic ground state only for the strained case.
}%
\label{fig3}%
\end{figure}

We can illustrate the physics of the recovery of magnetism under uniaxial strain on a simple toy model (Fig.~\ref{fig3}). Let us assume that the energy of the SDW in the unstrained state in the mean field theory (MFT) is a simple quadratic polynomial of its amplitude: $E(m)=-m^{2}+2m^{4}.$ The MFT ground state is then $m=1/2.$ If we take Gaussian spin fluctuations of $\xi=0.6$, then, per Eq.\ref{2}, these will destabilize the static order and stabilize a nonmagnetic ground state. Let us now assume that under strain, the quadratic coefficient got reduced by a factor of two (a reduced tendency towards magnetism), $E_{strain}(m)=m^{2}/2+2m^{4}.$ This would shift the MFT solution to $m=1/2\sqrt{2},$ and the magnetic stabilization energy will be reduced from $1/8$ ($E(m=1/2)$) to $1/32$ ($E_{strain}(m=1/2\sqrt{2})$). If we now assume that $\xi$ has also been reduced by a factor of two, from $0.6$ to $0.3,$ the SCR solution will still be magnetic, $m\approx0.42,$ $E_{strain}\approx-0.06.$.

\emph{Summary and discussions.}
Our DFT calculations show that, as expected,
the stress generally weakens the tendency toward magnetic order in Sr$_{2}$RuO$_{4}$.
That is to say, the magnetic ground state is less stable in the uniaxially strained
case than in the unstrained one on the mean-field level. However, at the same time,
the propensity of magnetic order to be destroyed by  quantum spin fluctuations
becomes  weaker and allows the establishment of long-range order in the
strained system. The competition between the two effects can be understood as follows:
The mean square amplitude of spin fluctuations $\langle M(\mathbf{q})^{2}\rangle$
decreases much faster with applied strain than static magnetic moment formation energy.
In the parlance of Moriya's SCR theory, the coefficient
$b$ in Eq. \ref{1} is negative, indicating the magnetic tendency, and becomes less negative with increasing the uniaxial strain. However, the strength of the spin fluctuation quantified by the parameter $\xi$, $also$ decreases with strain and, apparently, varies  faster than $b.$ As a result, up to some critical stress, determined experimentally as $\sim$ 0.8 GPa, $\tilde {b}=b+\frac{5}{3}c\xi^{2}$ remains positive and becomes negative afterward.

A corollary of this picture is that under further straining, the ordered magnetism will be suppressed again due to the further reduction of the magnetic tendency - a prediction that should be possible to verify
experimentally.

It is of note that the position of the MFT instability (or, equivalently,
of the maximum in spin susceptibility) shifts with strain from its unstrained
position of $\mathbf{q}=(0.29,0.29)$ toward $\mathbf{q}=(1/2,1/4),$
a new suggested nesting for strained case~\cite{Roemer2020}.
But, still, $\mathbf{q}_{SDW}$ remains
strongly incommensurate, $\mathbf{q\approx}(0.34,0.20),$ even at the strain
of several GPa, and this evolution does not reflect the changes in any types of nestings.
The origin of the deviation in $\mathbf{q}_{SDW}$, and its correlation with the
nesting vector is also a topic of further studies. While direct verification of this
prediction by neutron scattering is questionable, there might be
observable indirect manifestations.

Our investigations on the unusual emergence of magnetism in Sr$_{2}$RuO$_{4}$
can offer crucial insights into other unconventional superconductors where the pairing
mechanism is attributed to the spin-fluctuations.

\emph{Acknowledgments.} We thank Stuart Brown for fruitful discussions.
This work was supported by the National Research Foundation of Korea(NRF)
funded by the Ministry of Science and ICT (No. 2022M3H4A1A04074153).
B.K. also acknowledges support by NRF grant No. 2021R1C1C1007017 and KISTI
supercomputing Center (Project No. KSC-2021-CRE-0605).
B.K, S.K., and C.F is supported by Korea-Austria Joint Mobility Projects (NRF-2019K1A3A1A18116063).
I.I.M. acknowledges support from the U.S. Department of Energy through
the grant No. DE-SC0021089.

\end{document}